\documentclass[aps,twocolumn]{revtex4}
\usepackage{graphicx}

\begin{document}
\title{Analysis of self-thermalization dynamics in the Bose-Hubbard model \\ by using the pseudoclassical approach}
\author{Andrey R. Kolovsky}
\affiliation{Kirensky Institute of Physics, 660036, Krasnoyarsk, Russia} 
\affiliation{Siberian Federal University, 660041 Krasnoyarsk, Russia} 
\date{\today}

\begin{abstract}
We analyze the self-thermalization dynamics of the $M$-site Bose-Hubbard model in ring geometry using the single-particle density matrix calculated via the pseudoclassical approach. We show that a weak inter-particle interaction, which has a negligible effect on the thermal state given by the Bose-Einstein distribution, suffices to convert the integrable system of non-interacting bosons into a chaotic system. Next we consider two systems of weakly interacting bosons (i.e., two Bose-Hubbard systems) in different thermal states which are coupled by a point contact. We demonstrate that the chaotic nature of the Bose-Hubbard model ensures irreversible relaxation of both systems to global thermal equilibrium. Finally, substituting the point contact with a lattice of length $L\ll M$, we numerically calculate the quasi-stationary current of Bose particles across the lattice and show that its magnitude is consistent with the solution of the master equation for the boundary-driven $L$-site Bose-Hubbard system.
\end{abstract}
\maketitle

\section{Introduction}
\label{sec1}
The problem of self-thermalization in many-body systems of interacting particles can be traced back to the celebrated work by Fermi, Pasta, and Ulam \cite{Ferm55}, where the authors numerically studied the dynamics of interacting classical phonons. As is well known, they did not observe thermalization which at that time came as a big surprise. It took decades to realize that thermalization requires the presence of developed chaos \cite{Izra68,5} and, if it is the case, the system does thermalize \cite{Berm05,Chri25}. In quantum physics the problem of self-thermalization has been discussed since the beginning of the millennium when the main focus in the field of Quantum Chaos has shifted to  many-body systems.  Nowadays, the classical, quantum, and mixed quantum-classical approaches to the self-thermalization problem are an area of active theoretical \cite{Cass09,Polk11,Borg16,Rigo16,Beug14,Engl14,Schl16,Qiao25}, and experimental research, with the system of ultra-cold atoms in optical lattices being the most popular platform \cite{Trot12,Kauf16}.

In this work we study self-thermalization dynamics of the Bose-Hubbard  (BH) model -- one of the paradigm models for many-body classical and quantum chaos \cite{66,103,Paus21}. In relation to this particular system some aspects of thermalization are discussed in papers  \cite{Cass09,Beug14,Engl14,Schl16,Qiao25}. In contrast to the cited works, the new aspect is that we address not only the equipartition of populations of the individual sites but the genuine thermalization, where the system relaxes to the thermal state with well-defined values of temperature and chemical potential.
 
Along with the thermalization problem we also discuss the important problem of non-equilibrium statistical mechanics, namely, the stationary current of particles across the tight-binding chain connected at its edges to particle reservoirs with different chemical potentials. We approach this problem from first principles by modeling reservoirs as finite size BH systems. It is shown, in particular, that the occurrence of the stationary current requires the chaotic dynamics of reservoirs. We calculate the current numerically and compare results with the solution of the master equation for the boundary-driven BH model \cite{112}.

\section{Thermal density matrix}
\label{sec2}
\begin{figure}
\centering
\includegraphics[width=.5\textwidth]{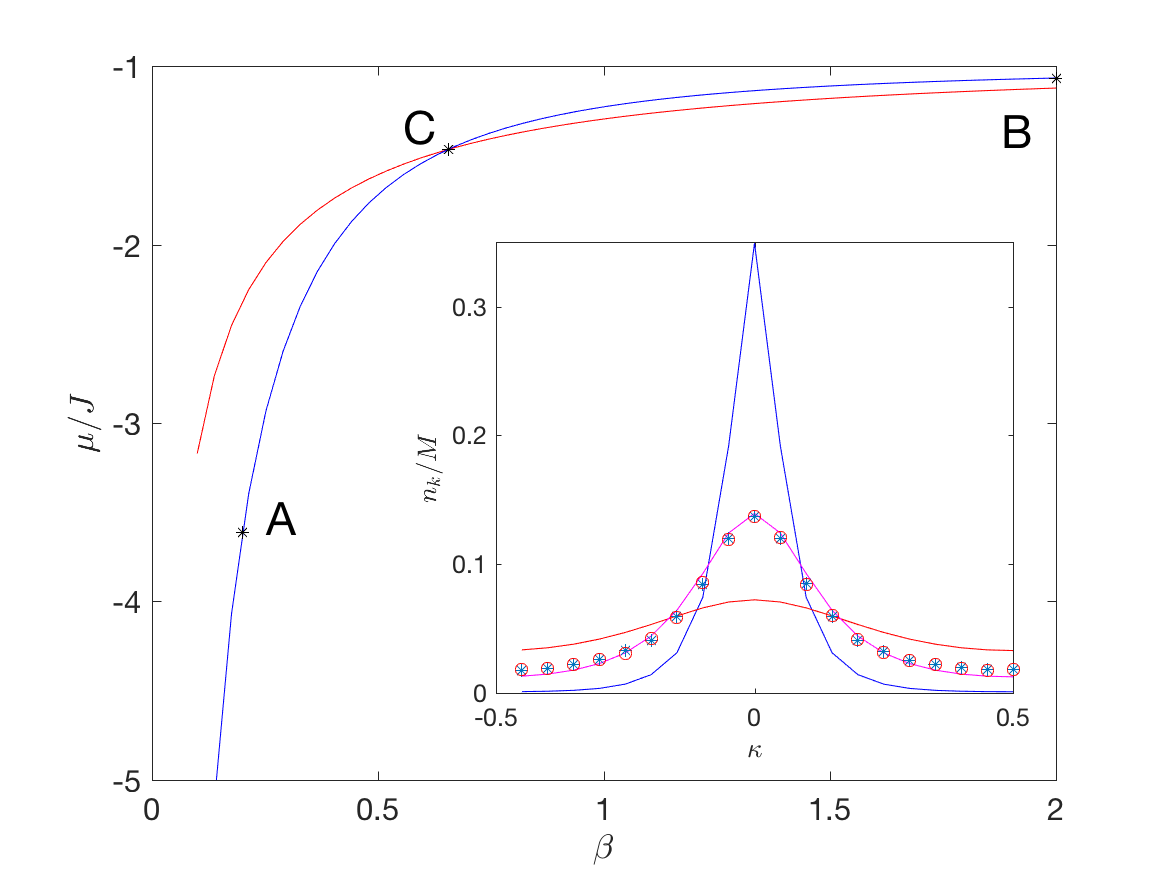}
\caption{Main panel: The line of constant density $\bar{n}=1$, blue dashed curve, and the line of constant energy $\bar{E}=-0.533$, solid read curve. Inset: Bose-Einstein distributions as the function of the quasimomentum $\kappa$ for the points $A$, $C$, and $B$, from bottom to top at $\kappa=0$. Asterisks and open circles show the result of equilibration dynamics for two coupled rings of the size $M=20$, see Sec.~\ref{sec3}.}
\label{fig00}
\end{figure}

The quantum Hamiltonian of the $M$-site BH model reads 
\begin{equation} \label{BH}
\widehat{\cal H}=
  -\frac{J}{2} \sum_{\ell=1}^M \left( \hat{a}^\dag_{\ell+1}\hat{a}_\ell +h.c.\right) +\frac{U}{2}\sum_{\ell=1}^M \hat{n}_\ell(\hat{n}_\ell-1) \;, 
 \end{equation}
where $\hat{n}_\ell=\hat{a}^\dagger_\ell\hat{a}_\ell$. It is parametrized by three independent  parameters -- the hopping matrix element $J$, the microscopic interaction constant $U$, and the mean particle density $\bar{n}=N/M$. Notice that  in the case of periodic boundary condition (the BH ring, in what follows) the Hamiltonian (\ref{BH}) can be rewritten in terms of the operators $\hat{b}_k=M^{-1/2} \sum_\ell \exp(i2\pi k\ell/M)\hat{a}_\ell$ , taking the form
\begin{eqnarray}\label{bloch}
\nonumber
\widehat{\cal H}= -J\sum_k \cos\left(\frac{2\pi k}{M}\right)\hat{b}^\dagger_k\hat{b}_k  \\+  \frac{U}{2L}\sum_{k_1,k_2,k_3,k_4} 
     \hat{b}^\dagger_{k_1}\hat{b}^\dagger_{k_2}\hat{b}_{k_3}\hat{b}_{k_4}\delta(k_1+k_2-k_3-k_4)  \;.
 \end{eqnarray}
These two representations of the BH Hamiltonian, which we shall refer to as the Wannier and the Bloch representations, help us to identify the two integrable limits $J=0$, $U\ne0$ and $U=0$, $J\ne0$ where the energy spectrum of the system can be calculated analytically. In the general case $J,U\ne0$ the BH model is known to be a chaotic system with energy spectrum obeying the Wigner-Dyson distribution \cite{66}. This result suggest  that  dynamics of the many-body density matrix ${\cal R}(t)$,  which  obeys the quantum Liouville equation
\begin{equation}\label{liuvil_qu}
\frac{{\rm d} {\cal R}}{{\rm d}t}=-i[\widehat{\cal H},{\cal R}]  \;,
\end{equation}
should show signatures of self-thermalization. All these make the BH model  a nice toy model for reviewing  different aspects of statistical physics from first principles. In particular,  according to the postulates of statistical physics, the single-particle density matrix (SPDM) of Bose particles in a lattice, 
\begin{equation}\label{spdm1}
\rho_{k,k'}(t)= {\rm Tr}[\hat{b}_k^\dagger \hat{b}_{k'}{\cal R}(t)] \;, 
\end{equation}
should relax to the steady-state density matrix
\begin{equation}\label{thermal}
\rho_{k,k'}=\delta_{k,k'} n_k \;,\quad n_k=\frac{1}{e^{\beta(E_k-\mu)}-1} \;, 
\end{equation}
where $E_k=-J\cos(2\pi k/M)$ are energies of the single-particle eigenstates and $\beta$ and $\mu$ are the inverse temperature and the chemical potential, respectively \cite{book}. These two parameters uniquely determine the mean particle density of Bose particles in a lattice, $\bar{n}=\sum_{k} n_k/M$, and the mean energy per particle or lattice site, $\bar{E}=\sum_{k} E_k n_k/M$. The other assertion is also true, namely, the particle density $\bar{n}$ and the mean energy $\bar{E}$ uniquely determines the system temperature $1/\beta$ and the chemical potential $\mu$. Notice that the latter is negative and cannot exceed  $-J$, see Fig.~\ref{fig00}. 

Let us now discuss the thermal density matrix (\ref{thermal}) from the viewpoint of the pseudoclassical  approach.  In what follows we prefer to work in the Wannier basis where the thermal density matrix is related to the matrix (\ref{thermal}) by the discrete Fourier transformation. 

For the finite size BH model the pseudoclassical  approach is based on the notion of the Husimi function  
$f_{qu}({\bf a},{\bf a}^*;t)=\langle{\bf a}| {\cal R}(t) |\bf{a}\rangle$,  where  $|\bf{a}\rangle$ are the coherent $SU(L)$ states \cite{Pere86,Trim08}. Similar to the truncated Wigner function approach, the pseudoclassical  approach amounts to approximating the governing equation for the Husimi function, which one obtains from the quantum Liouville equation (\ref{liuvil_qu}) \cite{Trim08}, by  the classical Liouville equation
\begin{equation}\label{liuvil_cl}
\frac{\partial f_{cl}}{\partial t}=\{H,f_{cl}\}  \;,\quad   f_{cl}({\bf a},{\bf a}^*;t=0)=f_{qu}({\bf a},{\bf a}^*;t=0) \;,
\end{equation}
where $H$ is the classical counterpart of the quantum Hamiltonian (\ref{BH}), 
\begin{equation}\label{BH_cl}
H= -\frac{J}{2}\sum_{l=1}^M (a^*_{l+1}a_l + c.c.) + \frac{g}{2}\sum_{l=1}^M |a_l|^4  \;, \quad g=U\bar{n} \;.
\end{equation}
Formally, the Hamiltonian (\ref{BH_cl}) follows from the quantum Hamiltonian by identifying the scaled annihilation and creation operators, $\hat{a}_\ell/\sqrt{\bar{n}}$ and $\hat{a}_\ell^\dagger/\sqrt{\bar{n}}$, with the pair of canonically conjugated variables $a_\ell$ and $a_\ell^*$  \cite{116}. Notice that $\sum_\ell |a_\ell |^2=M$ is the integral of motion of the classical BH model. The above scaling also implies the rescaled particle density $\bar{n}=1$.
 
Within the framework of the pseudoclassical approach elements of the SPDM $\rho_{\ell,\ell'}(t)= {\rm Tr}[\hat{a}_\ell^\dagger \hat{a}_{\ell'}{\cal R}(t)]$ are found as
\begin{equation}\label{spdm}
\rho_{\ell,\ell'}(t)= \int a_\ell^*a_{\ell'}  f_{cl}({\bf a},{\bf a}^*;t) {\rm d} {\bf a} {\rm d} {\bf a}^* \;,
\end{equation}
where the integral is taken over the whole phase space. Of course, calculating the multi-dimensional integral (\ref{spdm}) analytically is not an easy task. Fortunately, this integral can be evaluated by using Monte-Carlo simulations. 
The procedure goes as follows.  It is easy to prove  that at $t=0$ the classical distribution function can be sampled by the following ensemble of initial conditions
\begin{equation}\label{ensemble}
a_\ell=\frac{1}{\sqrt{M}}\sum_{k=1}^M \exp\left(\frac{2\pi k\ell}{M}\right)b_k \;,\quad b_k=\sqrt{n_k}e^{i\phi_k} \;,
\end{equation}
where $\phi_k$ are independent random phases and $n_k$ are given in Eq.~(\ref{thermal}). Although this ensemble is not unique it gives correct SPDM  Eq.~(\ref{thermal}). In what follows we shall call this ensemble of initial conditions by the quantum ensemble. Then, evolving these trajectories in time, the elements of the SPDM Eq.~(\ref{spdm}) are calculated as 
\begin{equation}\label{average}
\rho_{\ell,\ell'}=\langle a_\ell^*(t) a_{\ell'}(t) \rangle \;,
\end{equation}
where the angular brackets denote the average over the quantum ensemble of initial conditions.  Notice that, although the amplitudes $a_\ell(t)$ depend on time, the average (\ref{average}) does not depend on time if $g=0$.  This statement follows from the fact that for non-interacting bosons the pseudoclassical approach is exact and that for $g=0$ the matrix (\ref{thermal}) is a stationary state. We use this fact to control  convergence of the Monte-Carlo method. It appears that a few thousand  initial conditions are  sufficient to get good convergence. Further on we use 2048 trajectories. The accuracy of numerical simulations of ${\bf a}(t)$ is controlled by watching the trajectory energies, which are the same for every member of the quantum ensemble if $g=0$.
\begin{figure}
\centering
\includegraphics[width=.5\textwidth]{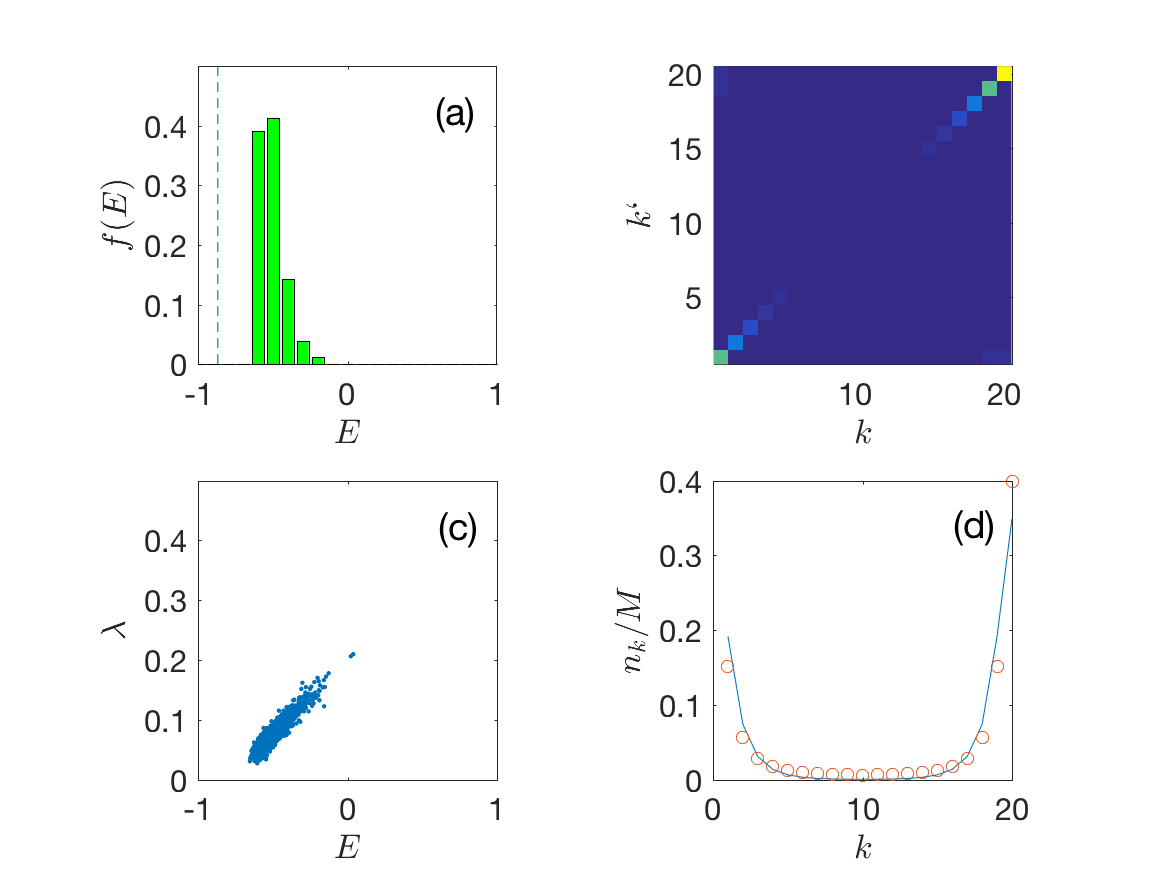}
\caption{Thermal density matrix for $\beta=2$, $\bar{n}=1$, and $g=0.4$: (a) distribution of 2048 trajectories from the quantum ensemble over the energy; (b) the stationary SPDM in the Bloch basis; (c) Lyapunov exponents of 2048 trajectories; (d) diagonal  matrix elements of the stationary SPDM, open circles, as compared to the Bose-Einstein distribution, solid line.}
\label{fig1}
\end{figure}
 
Is the matrix (\ref{thermal}) a steady state in the case of weakly interacting bosons?  When answering this question we focus on two cases: the low-temperature regime $\beta/J=2$ and the high-temperature regime $\beta/J=0.2$. In both cases the chemical potential  $\mu$ is adjusted to the value ensuring the rescaled particle density $\bar{n}=1$.  

For interacting bosons we evolve trajectories according to the nonlinear equation
\begin{equation} \label{eq_H}
i\dot{a}_\ell=\frac{\partial H}{\partial a^*_\ell} = -\frac{J}{2}\left(a_{\ell+1} +a_{\ell-1}\right) + g|a_\ell |^2 a_\ell 
\end{equation}
and we set $J=1$ from now on. We mention that, unlike the case $g=0$, now different trajectories belong to different energy shells. The distribution of trajectories over the energy for $g=0.4$ and $\beta=2$ is depicted in Fig.~\ref{fig1}(a), where the vertical line marks the trajectory energies for $g=0$. Let us also mention that for $g\ne0$ the ensemble Eq.~(\ref{ensemble}) is no longer an invariant set, i.e.,  the classical distribution function continuously evolves in time. This, however, does not prohibit the SPDM from relaxing to a steady state. We found that in the Bloch basis this new steady state  is again given by a diagonal matrix  with the diagonal elements $\rho_{k,k}$ only slightly deviating from  the Bose-Einstein distribution, see panels (b) and (d) in Fig.~\ref{fig1}. The observed relaxation of the SPDM to a steady state is a consequence of the coarse-graining  procedure Eq.~(\ref{spdm}) and  the chaotic dynamics of the system, which in terms of the distribution function implies the mixing dynamics of the latter (i.e., positive Kolmogorov-Sinai entropy) and positive Lyapunov exponents if discussed in terms of individual trajectories \cite{x}. For the considered $g=0.4$ the maximal Lyapunov exponents $\lambda_j$ of all trajectories from the quantum ensemble are plotted in panel (c).  It is seen that all trajectories have a positive Lyapunov exponent. Moreover, the mean exponent $\bar{\lambda}\equiv\langle\lambda_j\rangle$ monotonically increases with an increase in the interaction constant, see Appendix A. 
  
In principle, a small deviation of the SPDM of interacting bosons from the Bose-Einstein distribution can be quantified, see Appendix B. In what follows,  however, we shall neglect these deviations. Then the main conclusion of the present section is that the Bose-Einstein distribution, which refers to non-interacting bosons, also correctly captures the structure of the steady-state SPDM of weakly interacting bosons. However, the non-zero inter-particle interaction fundamentally changes the system dynamics, making it chaotic.

\section{Quantum manifestation of the mixing dynamics}
\label{secX}
As argued in the previous section, the origin of irreversible relaxation of the system SPDM to a steady state is the mixing dynamics of the classical distribution function. In this respect, it is interesting to discuss the quantum manifestation of this mixing dynamics in terms of the density matrix ${\cal R}(t)$, which is the quantum counterpart of the classical distribution function $f_{cl}(t)$. 

\begin{figure}
\centering
\includegraphics[width=.5\textwidth]{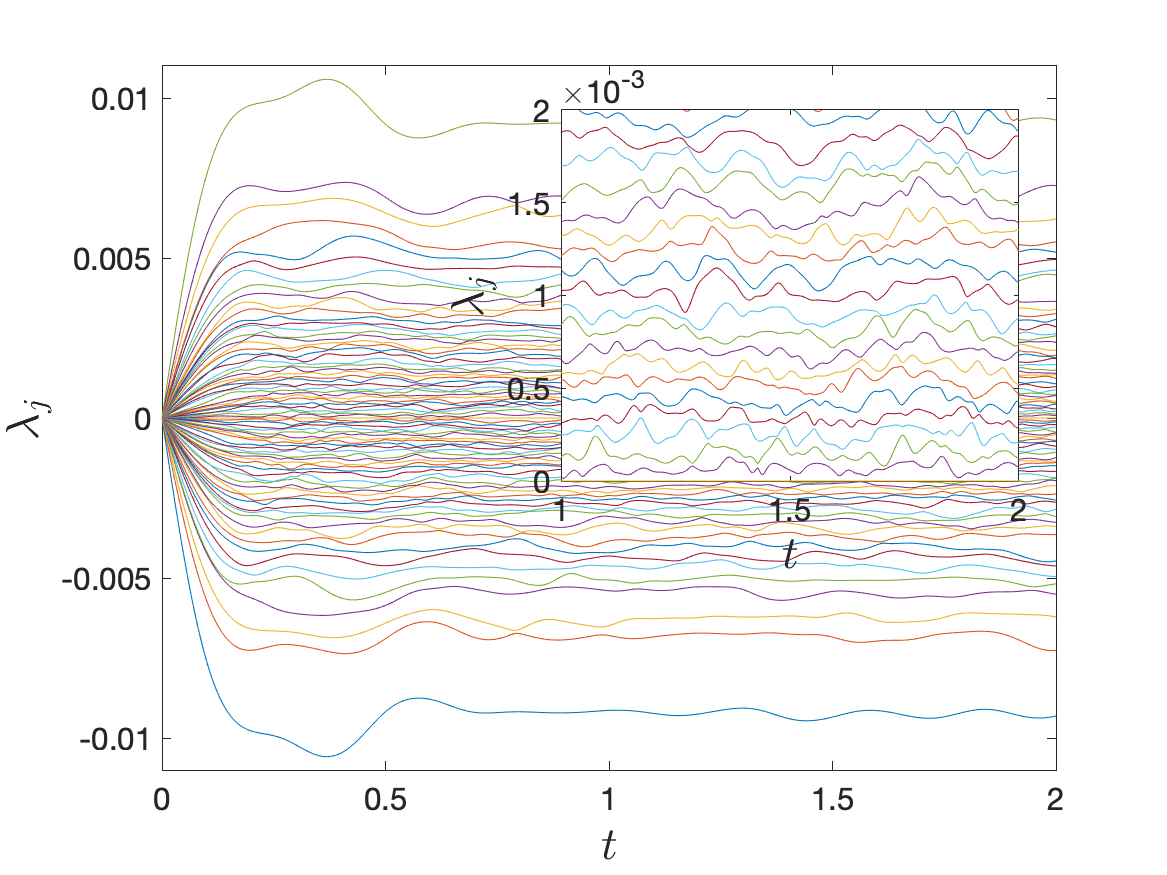}
\caption{Eigenvalues of the matrix $\Delta{\cal R}(t)$ as the function of time. System parameters are $U=0.4$, $N=M=6$, $\kappa=2\pi/M$ subspace. The inset zooms in the central part of the spectrum.}
\label{fig9}
\end{figure}

\begin{figure}[b]
\centering
\includegraphics[width=.5\textwidth]{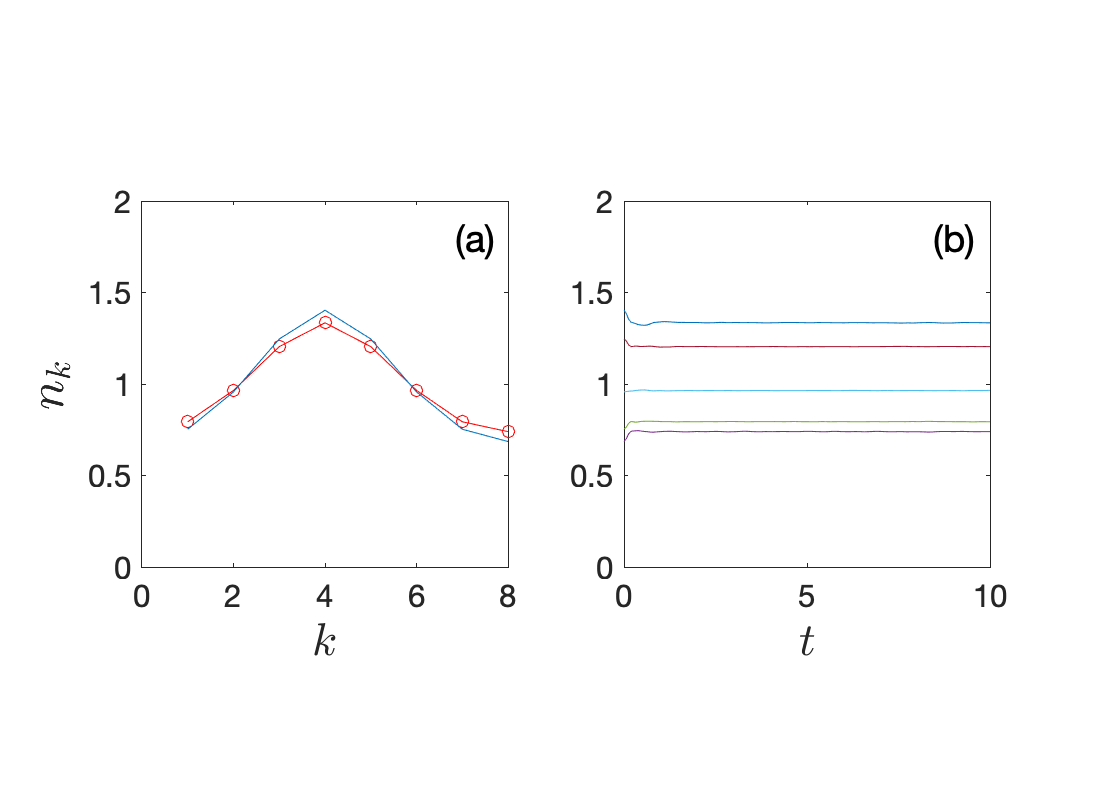}
\caption{Relaxation to the steady-state SPDM, right panel, and diagonal elements of the steady-state SPDM as compared to the Bose-Einstein distribution. Parameters are $\beta=0.2$, $U=0.4$, $N=M=8$, $\kappa=0$ subspace.}
\label{fig10}
\end{figure}

Let us analyze the solution of Eq.~(\ref{liuvil_qu}) for ${\cal R}(t=0)$ given by the thermal state 
\begin{equation}\label{total}
{\cal R}_0= \sum_{j=1}^{\cal N} P_j |{\bf n}_j\rangle\langle {\bf n}_j | \;.
\end{equation}
In Eq.~(\ref{total}) $|{\bf n}_j\rangle=|n_1^{(j)}, \ldots,n_k^{(j)}, \ldots,n_M^{(j)} \rangle$  are the quasimomentum Fock states and $P_j$ the weight coefficients determined according to postulates of statistical mechanics as $P_j\sim \prod_{k=1}^M p_k$ where $p_k$ is probability to find $n_k$ bosons in the $k$th quasimomentum state,
\begin{equation}\label{prob}
p_k=\left(1-e^{-\beta(E_k-\mu)}\right)e^{-\beta(E_k-\mu)n_k^{(j)}} \;.
\end{equation}
Since the matrix ${\cal R}_0$ is diagonal in the Bloch basis it is obviously the stationary solution of Eq.~(\ref{liuvil_qu}) if the interaction constant $U=0$. However, if $U\ne0$ the many-body density ${\cal R}(t)$ continuously evolves in time ant it is convenient to describe this evolution in terms of the matrix $\Delta{\cal R}(t)={\cal R}(t)-{\cal R}_0$. As an example, Fig.~\ref{fig9} shows eigenvalues of the matrix $\Delta{\cal R}(t)$ as the function of time for $U=0.4$, $N=M=6$, and the total quasimomentum $\kappa=2\pi/M$. (Notice that the total quasimomentum is the quantum integral of motion due to the presence of the $\delta$-function in the Hamiltonian (\ref{bloch}).) One sees in Fig.~\ref{fig9} the spaghetti-like level flow typical for quantum chaotic systems. Furthermore, the numerical analysis reveals the Wigner-Dyson level spacing statistics for any $t$ outside the time interval $0\le t < t^*\approx0.5$. This time interval corresponds to the transient regime where the system SPDM relaxes to the new steady state, see Fig.~\ref{fig10}(b), which is found to be very close to the thermal SPDM given by the Bose-Einstein distribution,  see Fig.~\ref{fig10}(a).

To conclude this section we mention that the chaotic nature of the BH Hamiltonian alone does not guarantee mixing  because the dynamics of the density matrix ${\cal R}(t)$ also depends on the initial condition. In other words, for some specific  ${\cal R}_0={\cal R}(t=0)$ thermalization may be absent. In what follows, however, we restrict ourselves to  ${\cal R}_0$ given by the Bose-Einstein distribution for non-interacting bosons, where we always reach the global thermalization. Also, from now on we shall employ the pseudoclassical approach exclusively. The main reason for this choice is that  the pseudoclassical approach allows us to consider large BH systems up to $M=100$ that, in turn, allows us to  address  a number of fundamental problems of non-equilibrium statistical physics.

\section{Equilibration dynamics}
\label{sec3}
Let us consider the two BH rings which are in two different thermal states. To be specific, we shall label the rings by the subindex ${\rm L}$, the left ring, and ${\rm R}$, the right ring.  At $t=0$ we connect them at the single site by using a reduced hopping matrix element $\epsilon<J$.  Since the dynamics of every system are chaotic, it is expected that they will relax to the new thermal state which is the same for both rings. 
\begin{figure}[b]
\centering
\includegraphics[width=.5\textwidth]{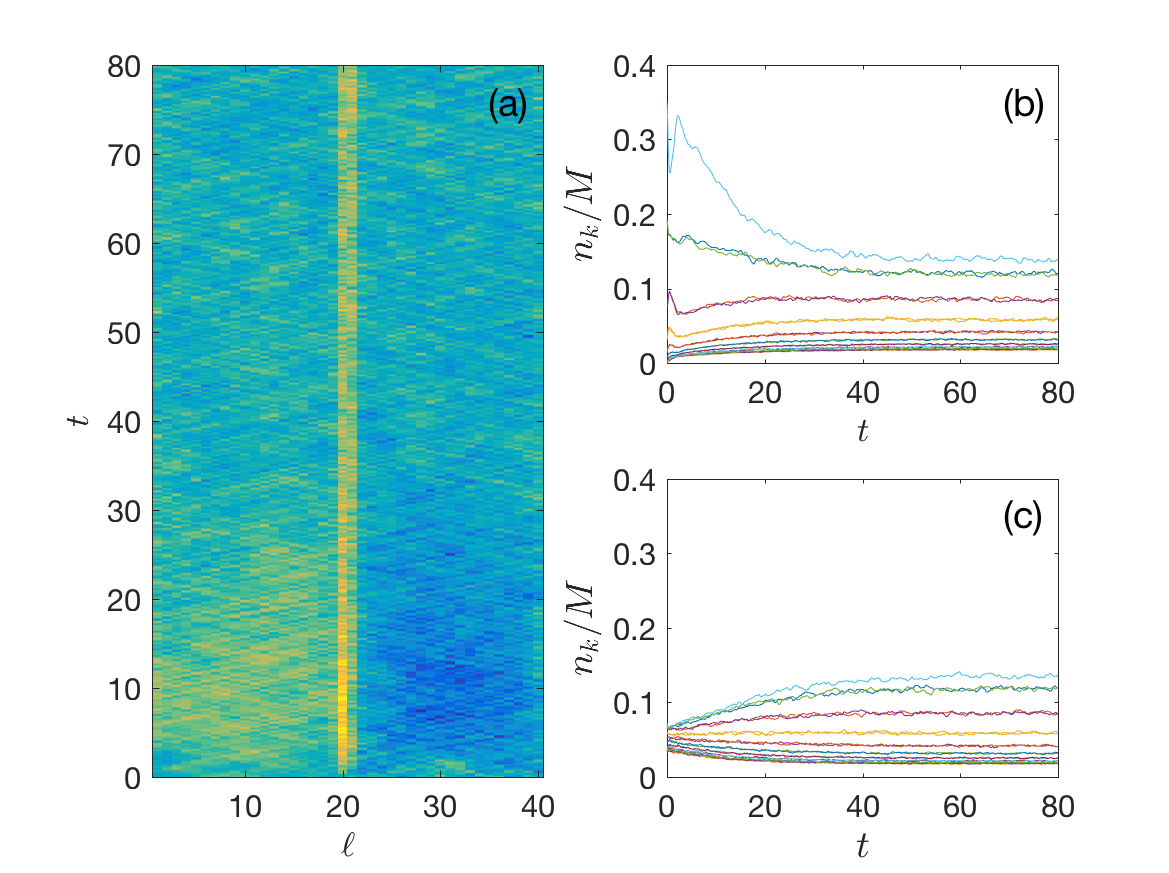}
\caption{Equilibration dynamics between two BH systems of the size $M=20$ which are  initially in the thermal states with   $(\beta,\mu)=(0.2,-3.63)$ and $(\beta,\mu)=(2,-1.07)$, see points $A$ and $B$ in Fig.~\ref{fig00}. The coupling constant  $\epsilon=0.25$ and the interaction constant $g=0.4$.}
\label{fig3}
\end{figure}

As an example we consider the case of equal particle densities $\bar{n}_{\rm L}=\bar{n}_{\rm R}=1$ but different temperatures $\beta_{\rm L}=2$ and $\beta_{\rm R}=0.2$  which corresponds to $\mu_{\rm L}\approx-1.07$ and $\mu_{\rm R}\approx-3.63$. Due to different temperatures the mean energies are also different, $\bar{E}_{\rm L}=-0.8713$ and $\bar{E}_{\rm R}=-0.1953$. When we bring the rings into contact, they can exchange particles and energy. This process is depicted  in Fig.~\ref{fig3}(a) showing populations of the lattices sites, where the interval $1\le\ell\le20$ refer to the left ring and the interval $21\le\ell\le40$ to the right ring. Right panels  in Fig.~\ref{fig3} show dynamics of diagonal elements of the ring SPDMs in the Bloch basis. It is seen in  Fig.~\ref{fig3} that in the course of time rings relax to the same steady state with the SPDM diagonal elements shown by open circles and asterisks in the inset in Fig.~\ref{fig00}. Remarkably, the latter perfectly match the Bose-Einstein distribution for $\bar{n}=1$ and $\bar{E}=( \bar{E}_{\rm L} + \bar{E}_{\rm R})/2=-0.533$, which is depicted by the solid magenta line in Fig.~\ref{fig00}.  We would like to stress that the discussed equilibration process involves the exchange of both energy and particles. Indeed, it is seen in Fig.~\ref{fig3}(a) that for short times hot bosons from the right ring enter the left ring, causing a temporal depletion  of the right  ring. Thus, during the  equilibration process the two BH systems generally do not adiabatically follow the blue line in Fig.~\ref{fig00} but come to the common thermal state (point $C$) only at the end of the equilibration passage.

\begin{figure}
\centering
\includegraphics[width=.5\textwidth]{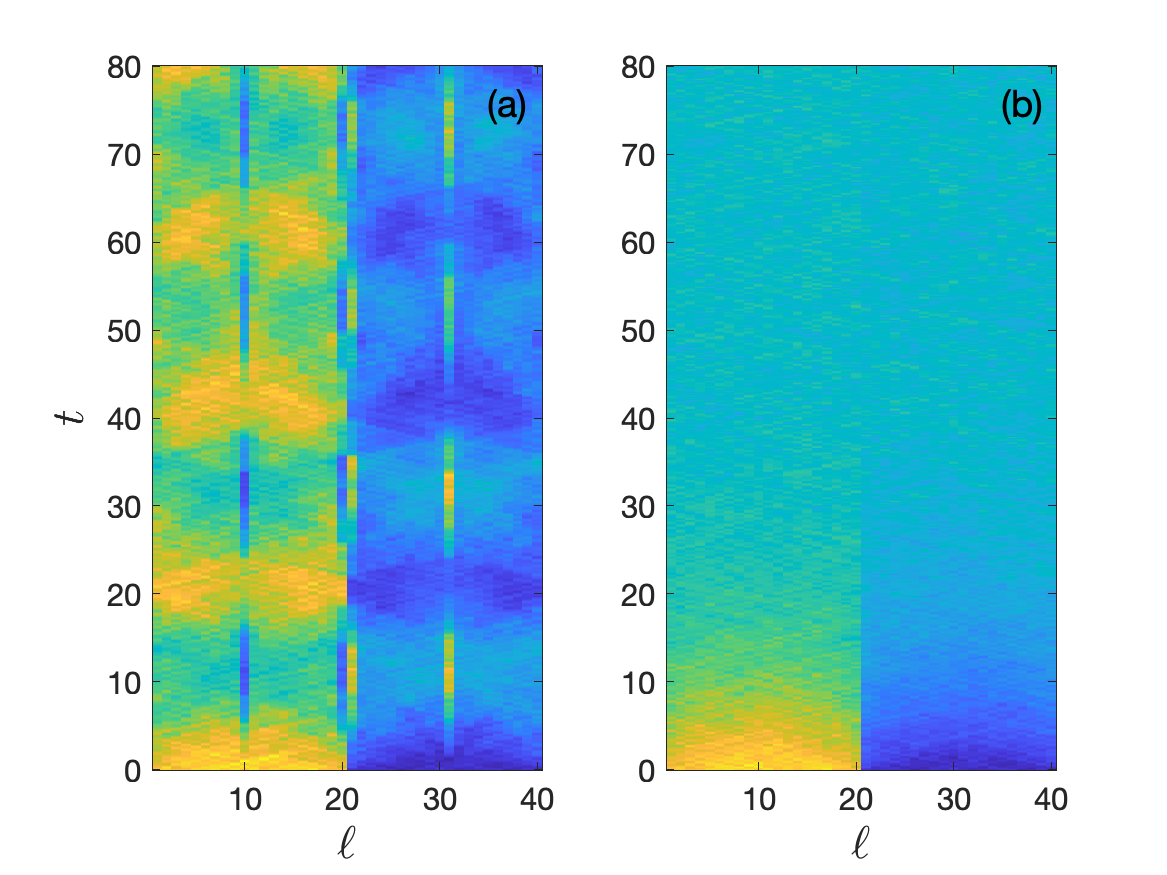}
\caption{Equilibration dynamics between two BH systems of size $M=20$ with initially different densities $\bar{n}_{\rm L}=1$ and $\bar{n}_{\rm R}=0.5$. The other parameters are $\beta_{\rm L}=\beta_{\rm R}=0.2$, $\epsilon=0.25$, and $g=0$, left panel, and $g=0.4$, right panel.}
\label{fig3b}
\end{figure}

Next we address the case of equal temperatures  $\beta_{\rm L}=\beta_{\rm R}=0.2$ but different densities $\bar{n}_{\rm L}=1$ and $\bar{n}_{\rm R}=0.5$, see Fig.~\ref{fig3b}. The left panel in this figure refers to the case of non-interacting particles $g=0$ where, as expected, we do not observe thermalization. This situation is in contrast to the case $g=0.4$ depicted in the right panel, where the system rapidly relaxes to the global thermal equilibrium with $\bar{n}_{\rm L}=\bar{n}_{\rm R}=0.75$.

\section{Non-equilibrium steady state}
\label{sec4}
In this section we discuss the stationary current of Bose particles across the lattice of the length $L$ connected at its edges to particle reservoirs with different chemical potentials.  The standard approach to this problem  of  non-equilibrium statistical mechanics consists of two steps. First, we eliminate reservoirs and derive the master equation for the reduced density  matrix ${\cal R}_s(t)={\rm Tr}_{\rm L,R}[{\cal R}(t)]$. For example, in the Markovian case this master equation has the form
\begin{equation} \label{master}
\frac{d {\cal R}_s}{d t}=-i[\widehat{\cal H}_s,{\cal R}_s] + \sum_{\rm i=L,R} {\cal L}_{\rm i}({\cal R}_s)  \;,
\end{equation}
where
\begin{eqnarray} \label{lind}
\nonumber
{\cal L}_{\rm i}({\cal R}_s)=-\frac{\Gamma_{\rm i}}{2} \left[(\bar{n}_{\rm i}+1)
\left(\hat{a}_{\rm i}^{\dagger}\hat{a}_{\rm i} {\cal R}_s-2\hat{a}_{\rm i}{\cal R}_s\hat{a}_{\rm i}^{\dagger}
+{\cal R }_s\hat{a}_{\rm i}^{\dagger}\hat{a}_{\rm i} \right)\right.  \\
+ \left.\bar{n}_{\rm i}
\left(\hat{a}_{\rm i}\hat{a}_{\rm i}^{\dagger}{\cal R}_s-2\hat{a}_{\rm i}^{\dagger}{\cal R}_s\hat{a}_{\rm i}
+{\cal R }\hat{a}_{\rm i}\hat{a}_{\rm i}^{\dagger} \right) \right] 
\end{eqnarray}
are the Lindblad relaxation operators and we use the convention  $\hat{a}_{\rm i=L}\equiv \hat{a}_1$ and $\hat{a}_{\rm i=R}\equiv \hat{a}_L$. Next,  we solve this master equation and find the stationary  reduced density matrix ${\cal R}_s={\cal R}_s(t\rightarrow\infty)$, which determines the stationary current across the chain. Remarkably, in the case of non-interacting bosons the stationary current can be found analytically \cite{112},  
\begin{equation} \label{current}
\bar{j}=\frac{J^2\Gamma}{J^2+\Gamma^2} \frac{\bar{n}_{\rm L}-\bar{n}_{\rm R}}{2} \;,
\end{equation}
where we set $\Gamma_{\rm L}=\Gamma_{\rm R}\equiv\Gamma$ for simplicity. The case of interacting bosons is more involved but it can also be studied, for example, by using the pseudoclassical approach \cite{116}. The main problem of the discussed two-step approach, however, is not  solving the master equation but controlling the approximations which were used to derive this equation.  Clearly, this is not possible without  going deeper into the reservoir structure, i.e., without having a microscopic or, at least, semi-microscopic \cite{123} model for the particle reservoir in hand.  The results of the previous section suggest that the BH model suits well to our purpose. 

We modify the two-ring setup of Sec.~\ref{sec3} and replace the point contact by a tight-binding chain of length $L$. It is also assumed that at $t=0$ the rings are in the high-temperature thermal states with the mean particle densities $\bar{n}_{\rm R}<\bar{n}_{\rm L}$. Of course, in numerical simulations we use rings of a finite size which may fully equilibrate quite rapidly. The way around this problem is to consider a small coupling constant $\epsilon$  and large $M$ where the equilibration time is much larger than other characteristic times of the system dynamics. Then, instead of the stationary current $\bar{j}$ we will have the quasi-stationary current  $\bar{j}(t)$. 
\begin{figure}
\centering
\includegraphics[width=.5\textwidth]{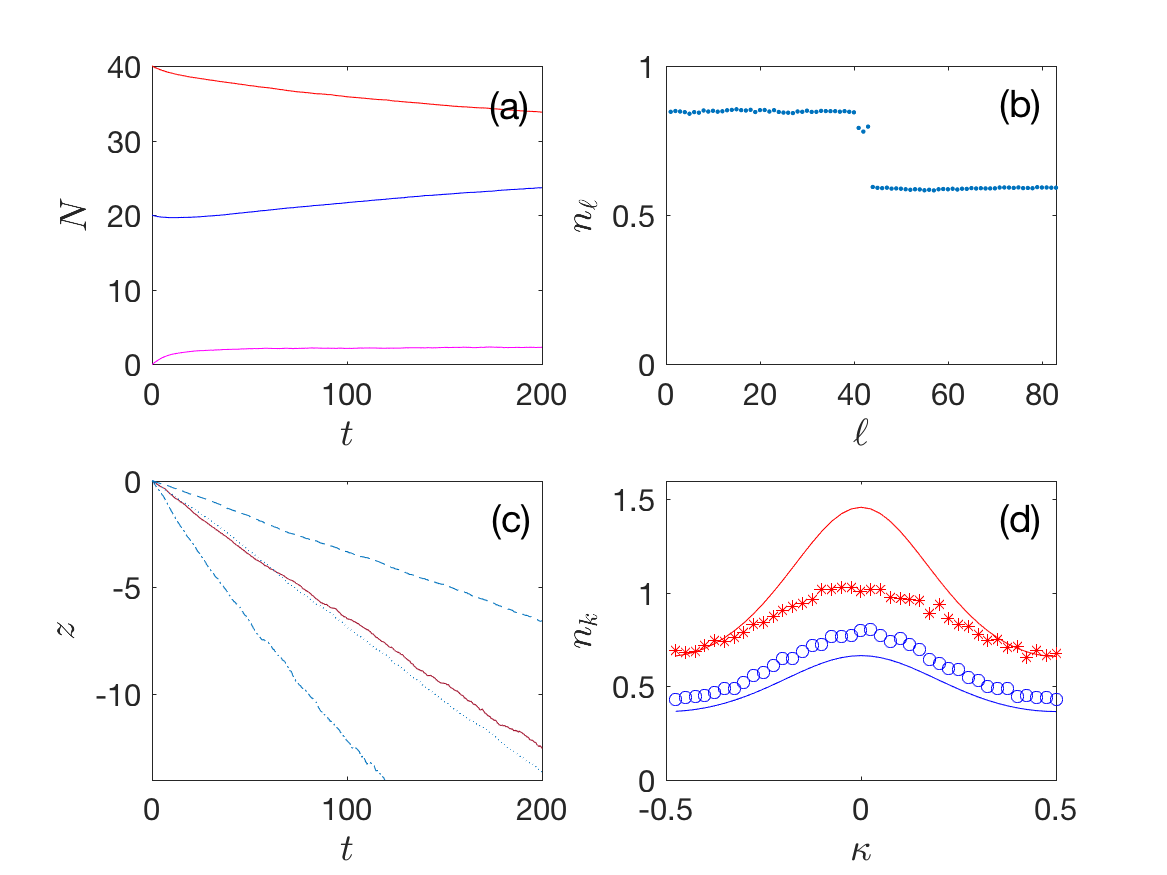}
\caption{ Equilibration dynamics for $\epsilon=0.1$, $M=40$, and $L=3$. The interaction constant for bosons in rings is $g=0.4$ while $g=0$ for bosons in the lattice. Initially rings are in the high-temperature thermal states with $\beta_{\rm L}=\beta_{\rm R}=0.2$,  and $\bar{n}_{\rm L}=1$ and $\bar{n}_{\rm R}=0.5$. Different panels show: (a) number of particles in rings and the lattice as a function of time; (b) populations of the individual sites at $t=200$;  (c) dynamics of the quantity (\ref{z}), red solid line. Additional dashed, dotted, and dash-dotted lines show $z(t)$ for $M=20$ and $\epsilon=0.1/\sqrt{2}$, $\epsilon=0.1$, and $\epsilon=0.1\sqrt{2}$; (d) diagonal elements of the ring SPDMs at $t=0$, solid lines, and $t=200$, markers.}
\label{fig6}
\end{figure}

Fig.~\ref{fig6}(a) shows dynamics of particle number in the left ring (upper line), the right ring (central line), and the lattice (lower line) for $\epsilon=0.1$, $M=40$,  and $L=3$. Initially the lattice is empty and the rings are in the thermal states with $\beta=0.2$ and $\bar{n}_{\rm L}=1$ ($\mu_{\rm L}\approx-3.61$) and   $\bar{n}_{\rm R}=0.5$ ($\mu_{\rm R}\approx-5.92$). Also, we set the interaction constant $g$ to zero for particles in the lattice.  It is seen in Fig.~\ref{fig6}(a) that after a short time required to populate the lattice we have a monotonic decrease/increase in the number of particles in  the left/right rings.  The main result is depicted in Fig.~\ref{fig6}(c) by the red solid line where we plot the quantity 
\begin{equation} \label{z}
z(t)=\frac{2}{M} \log\left(\frac{\Delta N(t)}{\Delta N(0)}\right) \;.
\end{equation}
The linear decrease of this quantity, $z=-Ct$, indicates that the quasi-stationary current in the lattice indeed obeys the relation
\begin{equation} \label{quasi}
\bar{j}(t)= C \frac{\bar{n}_{\rm L}(t)-\bar{n}_{\rm R}(t)}{2} \;, \quad
\bar{n}_{\rm i}(t)=\bar{n}_{\rm i}(0) \mp \int_0^t \bar{j}(t') {\rm d} t'  \;.
\end{equation}
We also depict in Fig.~\ref{fig6} the diagonal elements of the ring SPDMs, see panel (d), and occupations of all system sites, see panel (b),  where the three dots in the center refer to the lattice. 

Based on numerical data in  Fig.~\ref{fig6}(c) we find  the value of the relaxation constant $\Gamma$ entering Eqs.~(\ref{lind}) and (\ref{current}).  For $\epsilon=0.1$ it is $\Gamma\approx 0.07$. We repeated simulations for other values of the coupling constant $\epsilon$, see blue lines in Fig.~\ref{fig6}(c). These simulations confirm the expected quadratic dependence of  $\Gamma$ on $\epsilon$.  Next, following Ref.~\cite{123} we define the self-thermalization rate of the BH model as $\gamma=\epsilon^2/\Gamma$.  The obtained value $\gamma\approx 1/7$ is consistent with the validity condition of the Markovian master equation (\ref{master}), which requires that the self-thermalization time of particles in reservoirs has to be the shortest time scale of the problem.

\section{Concluding remarks}
First of all, we comment on our choice for the value of the macroscopic interaction constant $g=0.4$. This choice is dictated by the  two requirements. Namely,  to use textbook statistical mechanics the interaction constant has to be as small as possible. However, to have self-thermalization we need developed chaos that imposes the condition $g>g_{cr}$.  The chosen $g=0.4$ appears to be a good compromise between the above requirements.

The second remark concerns the self-thermalization rate $\gamma$ of the BH model. We stress that we distinguish between equilibration and self-thermalization times. The former depends on the system size  and typically implies that at $t=0$ the initial state of the system strongly deviates from the thermal state. On the contrary, the self-thermalization time is independent of the system size and indicates how quickly the SPDM of the BH system comes back to the  thermal state if we slightly perturb this state. The rate $\gamma$ also determines the decay rate $\gamma_c\approx\gamma$ of the reservoir autocorrelation functions which enters the master equation for the reduced density matrix of the carriers in the lattice. It is reasonable to assume that $\gamma$ is determined by the mean Lyapunov exponent of the BH model which, in its turn, is determined by the value of the interaction constant $g$. We find that $\gamma$ indeed increases as $g$ is increased, however, the exact functional dependence of $\gamma=\gamma(g)$ remains an open problem.

\vspace{0.5cm}
The author acknowledges financial support from state contract No FWES-2024-0003.

\section{Appendix A}

\begin{figure}
\centering
\includegraphics[width=.5\textwidth]{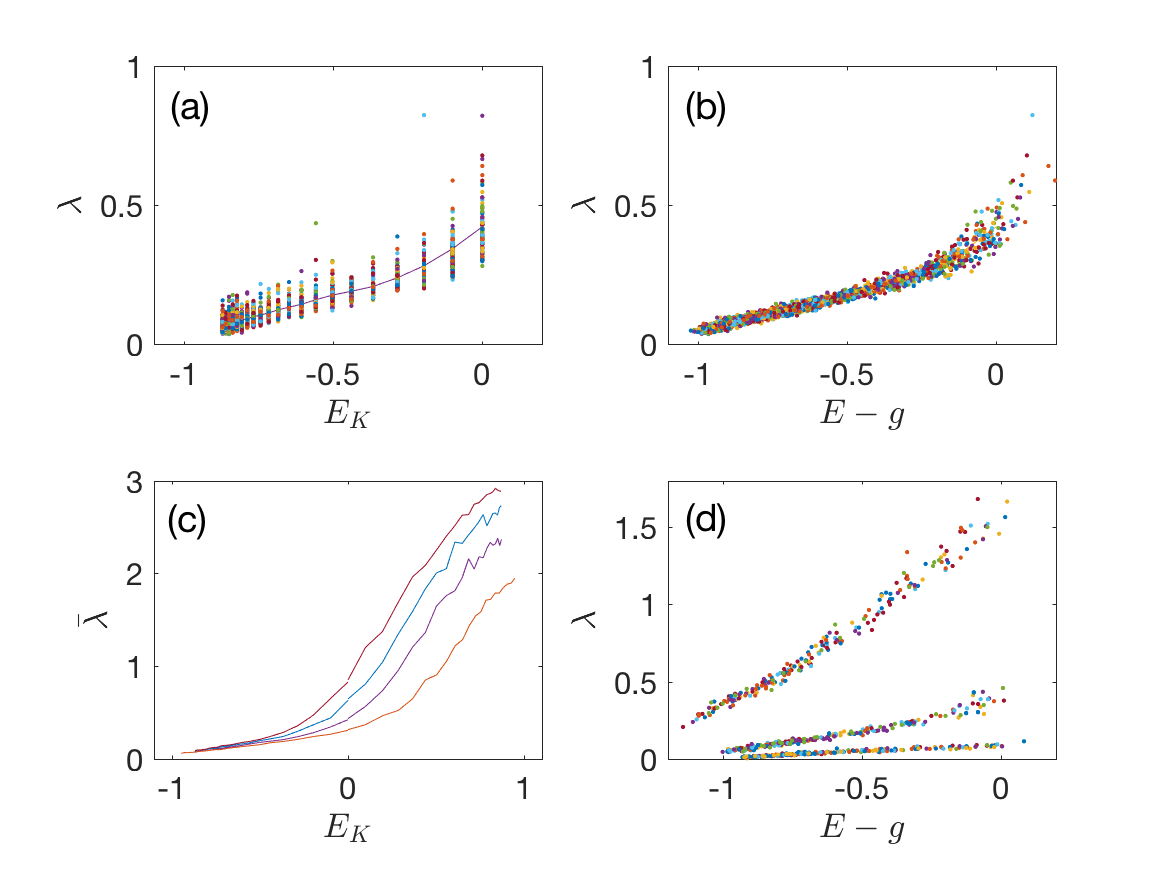}
\caption{(a) Lyapunov exponents for $g=0.4$, $\bar{n}=1$, and  $M=20$ as the function of the mean kinetic energy and (b) as the function of the total energy; (c) the mean Lyapunov exponent for $g=0.4$, $\bar{n}=1$, and  $M=10,20,40,80$, from bottom to top,  as the function of the mean kinetic energy (average over 100 initial conditions); (d) Lyapunov exponents for $\bar{n}=1$, $M=20$, and $g=0.2,0.4,0.8$, from bottom to top,  as the function of the total energy.}
\label{fig7}
\end{figure}

Here we collect results of our studies on the Lyapunov exponent of the classical BH model. We calculate the Lyapunov exponent  of a given trajectory by evolving the distance to its neighboring trajectory in tangential space. Fig.~\ref{fig7}(a) shows Lyapunov exponents for $g=0.4$, $\bar{n}=1$, and  equidistant  values of the inverse temperature in the interval $0\le\beta\le2$, which correspond to different discrete values of the kinetic energy $E_K(\beta)=\sum_k E_k n_k(\beta)/M$ in the interval $-J<E_K\le0$. Notice that trajectories are always taken from the quantum ensemble Eq.~(\ref{ensemble}).  Additionally, Fig.~\ref{fig7}(b) shows the same data yet the Lyapunov exponents are plotted as the function of the total energy $E=E_K+E_P$.  It is seen, that the Lyapunov exponents grow with the trajectory energy or, what is essentially the same, with the temperature.

We also studied the dependence of $\lambda$ on the interaction constant $g$, see  Fig.~\ref{fig7}(d),  and the system size $M$, see Fig.~\ref{fig7}(c), where four curves refer to $M=10,20,40,80$, from bottom to top. Notice that in Fig.~\ref{fig7}(c) we extend the analysis onto the case of negative temperatures. Here the term `negative temperature' means that the distribution function for the populations of the Bloch states is shifted by one-half of the Brillouin zone, i.e., in the limit $\beta\rightarrow-\infty$  only the Bloch state with the quasimomentum $\kappa=\pi$ is populated.

It is well-known that for $\kappa=\pi$ (in the general case, for $|\kappa|>\pi/2$) the plane-wave solution of the non-linear Schr\"odinger equation,
\begin{equation} 
a_\ell(t) = \exp[i\kappa\ell +iJ\cos(\kappa)t +igt] \;,
\end{equation}
exhibits modulation instability where the eigenmode with $\kappa=\pi$ decay into nearby modes which, in turn, exponentially grow in time, $|b_{\pi\pm q}(t)|^2 \sim \exp[\nu(q)t]$, with the increment \cite{5} 
\begin{equation} 
\nu(q)=|q|\sqrt{2g/J-q^2} \;. 
\end{equation}
This increment gives the upper boundary for the maximal Lyapunov exponent. The existence of the upper boundary implies that the mean Lyapunov exponent $\bar{\lambda}=\langle \lambda(E_K) \rangle$ is finite in the thermodynamic limit $M\rightarrow\infty$. However, as it is seen Fig.~\ref{fig7}(c), the convergence  is not uniform with respect to the trajectory energy.

\section{Appendix B}

\begin{figure}
\centering
\includegraphics[width=.5\textwidth]{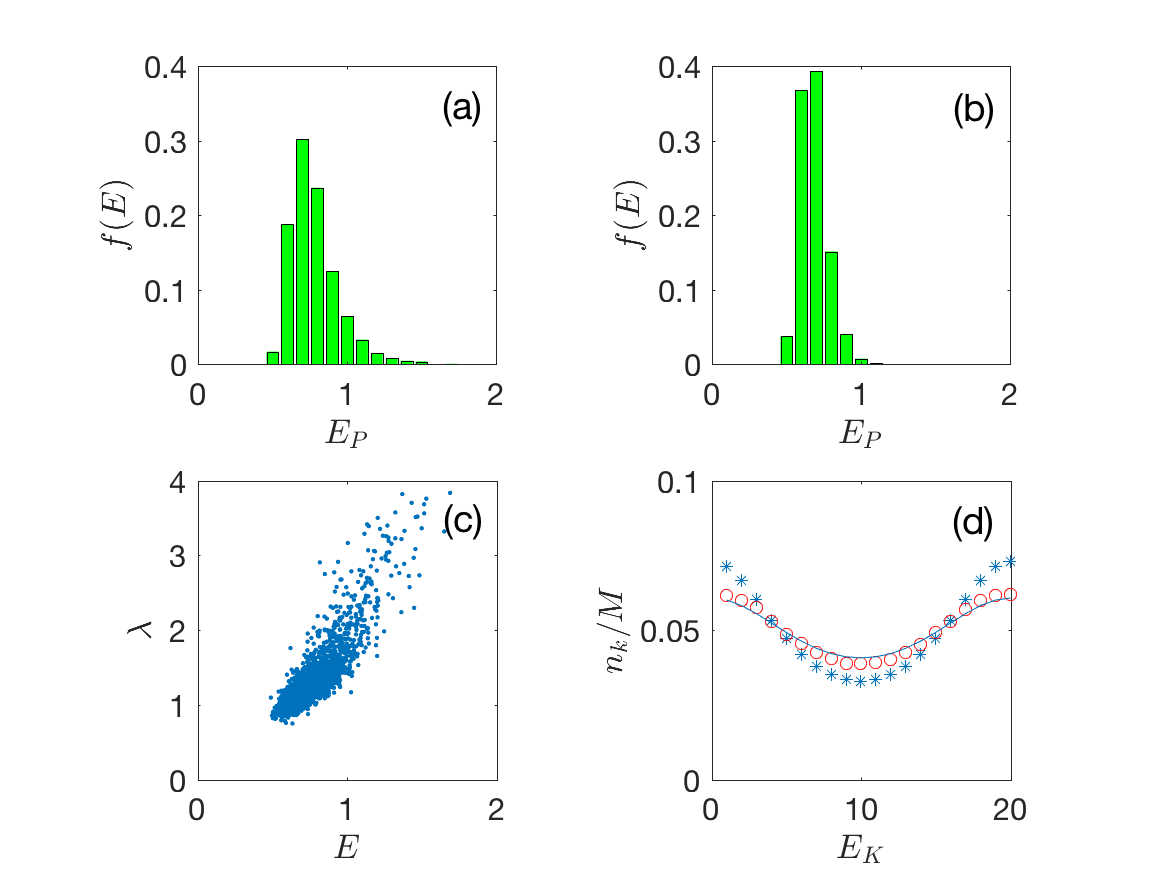}
\caption{Steady state SPDM for $g=0.8$: (a)-(b) potential energy of trajectories from the quantum ensemble at $t=0$ and $t$ much larger than the self-thermalization  time; (c) Lyapunov exponents; (d) diagonal  matrix elements of the stationary SPDM, open circles, as compared to the  initial  Bose-Einstein distribution with $\beta=0.2$, $\mu=-3.6125$ ($\bar{n}=1$), asterisks,  and  the Bose-Einstein distribution with adjusted $\beta=0.0980$ and $\mu=-7.1460$  ($\bar{n}=1$), solid line.}
\label{fig8}
\end{figure}

To have a reference point we repeated calculations of Sec.~\ref{sec2} for $\beta=0.2$, $\mu=-3.6125$ ($\bar{n}=1$), and a twice larger value of the interaction constant $g=0.8$, that leads to a larger deviation of the new steady-state SPDM from the Bose-Einstein distribution, see Fig.~\ref{fig8}(d). In order to qualify this deviation, we plot in the
upper panels in  Fig.~\ref{fig8} distributions of the potential energy $E_P$,
\begin{equation} 
E_P=\frac{g}{2}\sum_{\ell=1}^M |c_\ell |^4 
\end{equation}
for trajectories from the quantum ensemble for $t=0$ and $t\rightarrow\infty$.  These distributions give the following values for the mean potential energy per site -- $\bar{E}_P(0)=0.7798$ and $\bar{E}_P(\infty)=0.6825$. Thus, during the self-thermalization process  a part of the potential energy is transformed into kinetic energy. Then, if we identify the mean kinetic energy of bosons with the temperature, the diagonal elements of the SPDM should be compared not with the Bose-Einstein distribution for $\beta=0.2$ but with that for higher temperature which corresponds to  the new kinetic energy. This distribution is depicted in Fig.~\ref{fig8}(d) by the solid line and corresponds to $\beta=0.0980$ and $\mu=-7.1460$  ($\bar{n}=1$). A good agreement is noticed. 


\end{document}